\documentstyle [12pt] {article}
\title{ \Large \bf  Viscosity Prescriptions
in Accretion Disks with Shock Waves }
\author{Sandip K. Chakrabarti\\
{\small Tata Institute Of Fundamental Research, Bombay, 400005 INDIA}\\
and\\
Diego Molteni\\
{\small University of Palermo, Via Archirafi 36, 90123 Palermo, ITALY}\\}
\begin{document}
\baselineskip 22pt
\maketitle
\begin{abstract}

\baselineskip 22pt

We study the evolution of viscous, isothermal, rotating, thin, axisymmetric
accretion disks around a compact object using Smoothed Particle
Hydrodynamics. We emphasize the effects
of different choices of viscosity prescriptions on the evolution
of angular momentum as well as other physical quantities of the disk.
We show that a flow with the Shakura-Sunyaev viscosity prescription
may produce only {\it shear} shock, where the angular momentum
changes significantly across the shock. We present and study the effects of
other prescriptions of viscosity which render the angular momentum continuous
across the shock waves. In general, it is observed that for flows
with a small viscosity, the shocks are weaker, form farther away and
are wider as compared to the shocks in inviscid flows. If the viscosity
is high, shocks do not form at all. The flow remains subsonic and Keplerian
throughout the disk and becomes supersonic only very close to the horizon.

\end{abstract}

\newpage

\noindent {\bf 1. Introduction}

A large number of numerical simulation results present in the
literature suggest that shocks may form in adiabatic accretion disks
(Hawley, Smarr \& Wilson 1984; Chakrabarti \& Molteni 1993;
Molteni, Lanzafame \& Chakrabarti, 1994) and in inviscid disks
with radiative transfer (Molteni, 1994). These shock
solutions were obtained for inviscid flows and agree with the theoretical
predictions (Chakrabarti, 1989, 1990a) reasonably well. Chakrabarti (1990b,
hereafter SKC90b) discussed effects of viscosity
on the shock formation in isothermal accretion disks around black holes.
The general conclusions were that the stable shock ($X_{s3}$ in the notation of
SKC90b) is weaker and forms farther away as the viscosity parameter is
increased. When the viscosity is very high, shocks
do not form at all. Instead, the flow remains subsonic and Keplerian
throughout the disk and becomes supersonic close to the black hole
after passing through the inner sonic point. In the present paper,
we shall study the numerical evolution of the viscous isothermal disks with
a particular emphasis on the nature of shocks in flows close to a black
hole. We show that the solution strongly depends upon
the viscosity prescription. In particular, we show that if the transport of
angular momentum follows the Shakura-Sunyaev (1973) $\alpha$
viscosity (hereafter SS$\alpha$) prescription, a shear shock must
form, where a jump of angular momentum take place. We present other
prescriptions where angular momentum is continuous across the shock
waves and for high viscosity yields shock-free Keplerian disks exactly as
the Shakura-Sunyaev prescription. These prescriptions, therefore,
could possibly be more suitable for disks which include shock waves
as well as for shock free Keplerian disks.

The plan of our paper is the following: In the next Section, we
show how different viscosity prescriptions might affect the nature
of shock transitions. For simplicity of arguments we assume only the {\it
isothermal} flows. In Section 3, we present a test of our code
to show that with a significant viscosity, the disk does become Keplerian.
Subsequently, we present a few numerical solutions of the
isothermal flows with various viscosity prescriptions and compare these
results. Finally, in Section 4, we summarize our results and draw conclusions.

\noindent {\bf 2. Model Equations with Different Viscosity Prescriptions}

Consider a thin, isothermal, axisymmetric, accretion flow onto a compact
object. The radial momentum equation is:
$$
\frac{\partial v_r}{\partial t} + v_r \frac{\partial v_r}{\partial r}
+ \frac{1}{\Sigma}\frac{\partial W}{\partial r}-\frac{\lambda^2}{r^3} +
\frac{\partial \Phi}{\partial r}=0 .
\eqno{(1a)}
$$
The continuity equation is given by,
$$
\frac{\partial\Sigma}{\partial t}
+\frac{1}{r}\frac{\partial}{\partial r} (\Sigma r v_r)=0
\eqno{(1b)}
$$
The azimuthal momentum equation is given by,
$$
\frac{\partial \lambda}{\partial t} + v_r \frac {\partial \lambda}{
\partial r } = \frac{1}{r \Sigma} \frac{\partial}{\partial r} (r^2 W_{r \phi})
\eqno{(1c)}
$$

Here, $v_r$ and $\lambda$ are the radial velocity and the azimuthal
angular momentum respectively, and $W$, $\Sigma$ and $W_{r \phi}$
are the pressure, density  and the $r\phi$
component of the viscous stress tensor respectively.
$\Phi(r,\theta)$ is the gravitational potential of the central
object. This may be any of the Newtonian or pseudo-Newtonian
potentials. For example, $\Phi=-1/2(r-1)$ (Paczy\'nski \& Wiita, 1980)
could be used for flows around a Schwarzschild black hole
or a more complex form (Chakrabarti \& Khanna 1992) for flows around a
Kerr black hole. All distances, velocities and time scales are measured
in units of $2GM/c^2$, $c$ and $2GM/c^3$ respectively.

If we consider the solution to be steady, the time derivatives of the
above equations vanish and we get the following conservation equations,

\noindent (a) Conservation of energy:
$$
{\cal E}= \frac{1}{2} v_r^2 + K^2 log(\Sigma) + \frac{1}{2} \frac
{\lambda^2}{r^2} + \Phi
\eqno{(2a)}
$$
\noindent (b) Conservation of baryons,
$$
{\dot M}= \Sigma v_r r
\eqno{(2b)}
$$
and,\\
\noindent (c) Conservation of the angular momentum
$$
{\dot M} (\lambda-\lambda_{e}) = -r^2 W_{r \phi}
\eqno{(2c)}
$$
Here, we have used the polytropic equation of state $W= const. \ \Sigma$
appropriate for isothermal flows and $K=(W /\Sigma)^{1/2} $ is the
isothermal sound speed. Here, $\lambda_e$ is the angular momentum at the
inner edge of the disk.

A black hole accretion is necessarily transonic (Chakrabarti 1989, 1990ab).
An inviscid flow may be Bondi-like, i.e., simply passes through
a sonic horizon and falls onto a black hole supersonically. However,
as Liang and Thomson (1980) pointed out, the flow may have more than
one saddle type sonic points. Subsequently, is it shown that standing shocks
may be developed in rotating winds and accretion (Habbal \& Tsinganos 1983,
Ferrari et al. 1985, Fukue 1987, Chakrabarti 1989, 1990ab).
In particular, Chakrabarti (1990b) showed that the topological
properties of the phase space behaviour of the flow
depend strongly upon the angular momentum distribution and therefore
on the nature of viscosity.

In an inviscid, axisymmetric flow ($W_{r \phi}= 0$),
the characteristics (such as the location, strength, etc.)
of the shock is determined by the shock conditions
(Chakrabarti, 1990ab),\\

\noindent (a) Temperature of the flow is constant across the shock:
$$
K_-=K_+
\eqno{(3a)}
$$
\noindent (b) Baryon flux is conserved,
$$
{\dot M}_- = {\dot M}_+
\eqno{(3b)}
$$
\noindent (c) The total pressure is balanced,
$$
W_-+\Sigma_- v_{r-}^2 = W_++\Sigma_+ v_{r+}^2
\eqno{(3c)}
$$
and,\\
\noindent (d) Angular momentum  flux is conserved,
$$
\lambda_- = \lambda_+
\eqno{(3d)}
$$
Here, $-$ and $+$ signs represent quantities in the pre-shock and
post-shock flows respectively.

In the presence of viscosity, the angular momentum is transported
following Equation (2c).
In the following subsections, we discuss how the transport
process depends on the viscosity prescriptions.

\noindent {\large 2a. Shakura-Sunyaev Viscosity Prescription}

Let us first consider the effects of
SS$\alpha$ prescription. In this case, we have $W_{r\phi} = \alpha_{s} W \frac
{r}{\Omega_K}\frac{d\Omega}{dr}$,
and the angular momentum distribution is obtained from (SKC90b),
$$
\lambda - \lambda_e = \frac{\alpha_{s} r^3 W }{\Omega_K \dot M}\frac{d\Omega}
{dr} = \frac {\alpha_{s} r K}{M\Omega_K}\frac{d\Omega}{dr}
\eqno{(4)}
$$
where, $M=v_r/K$ is ths Mach number of the flow, and $\lambda_e$ is the
angular momentum at the inner edge of the disk and $\Omega_K$ is the
local Keplerian angular velocity. The subscript $s$ of
$\alpha$ is to indicate the Shakura-Sunyaev $\alpha$ parameter.
It is evident that at the shock, since $M$ varies from supersonic ($M_- >1$)
to subsonic ($M_+ <1$), the rate of transport of angular
momentum would be quite different on both sides of the shock.
In particular, in the post shock region, the angular momentum
would be `piling' up as the preshock flow is unable to transport
it away efficiently. Thus, the angular momentum must be
discontinuous and $\lambda_+ $ must be higher compared to $\lambda_-$.
Thus the shock formed would be of 'shear-type'.

\noindent {\large 2b. Continuous Angular Momentum Prescriptions}

In the case where the angular momentum is assumed to be continuous across the
shock wave, which is reasonable for any axisymmetric flows away from
narrow boundaries, one cannot use SS$\alpha$ prescription. Continuity of
$\lambda$ implies continuity of the viscous stress (cf. Equation 2c) and one
notes that at the shock (assumed here to be of infinitesimal width),
$$
W_{r \phi - } = W_{r \phi +}
\eqno{(5)}
$$
In the usual form:
$$
W_{r \phi} = \nu \Sigma r \frac{\partial \Omega}{\partial r}
\eqno{(6)}
$$
Here, $\nu$ is the kinematic viscosity coefficient, and $\Omega =\lambda/r^2$
is the local angular velocity. Since $\lambda$ is assumed to be
smooth and continuous, $\Omega$ is also a smooth and continuous
function. One way to achieve the continuity of $\lambda$ across the shock is to
define kinematic coefficient,
$$
\nu_p = \frac{\alpha_p (K^2 + v_r^2)}{\Omega_K}
\eqno{(7a)}
$$
so that the viscous stress is,
$$
W_{r \phi} =\frac {\alpha_p (K^2 + v_r^2) \Sigma r }{\Omega_k}
\frac{\partial \Omega}{\partial r}.
\eqno{(7b)}
$$
In this case, the pressure balance condition (Eqn. 3c)
ensures the continuity of $W_{r\ phi}$. This prescription will be
referred to as the `pressure balanced viscosity prescription'. The
subscript $p$ distinguishes this prescription from other ones.

One could instead use the balance of the mass flux in defining
the kinematic viscosity,
$$
\nu_m= \frac{\alpha_m v_r}{\Omega_K}
\eqno{(7c)}
$$
so that the viscous stress becomes,
$$
W_{r \phi} =\alpha_m |v_r| \Sigma  r \frac{\partial \Omega}{\partial r} .
\eqno{(7d)}
$$
The continuity of the viscous stress follows directly from the
conservation of the baryon flux (Eqn. 3b). This prescription will be
referred to as `mass flux balanced viscosity prescription'. The subscript
$m$ distinguishes this prescription from other ones.

\noindent{\large 2c. Prescription with Flux-limited Diffusion }

Recently, an important issue is brought up that in the boundary layer of
a slowly rotating star the SS$\alpha$ prescription implies that
the radial flow has to be supersonic and therefore would loss
the causal contact with the star surface (Narayan, 1992).
In this situation, the angular momentum is transported rather rapidly
and $\frac{\partial \Omega}{\partial r}$ cannot be a constant.
To remedy the problem of causality, a modified form of
the kinematic viscosity coefficient was prescribed:
$$
\nu_d = \alpha_d K^2 (1-v_r^2/v_t^2)^2 /\Omega_K
\eqno{(8)}
$$
Here, $v_t \sim \beta a$, (with $\beta$ of the order unity) is the turbulent
speed with which angular momentum
is transported fastest. This important modification  brings the disk
solution in the causal contact with the star by enforcing a subsonic
accretion. The subscript $d$ distinguishes this prescription from other ones.

In the case of black hole accretion this specific problem does not arise
since the inner boundary condition dictates that the flow {\it has to be}
supersonic. Secondly, as in the case of disks in stellar systems,
angular momentum transport is not always due to hydrodynamical
processes. Radiative, magnetic and other means of transport processes
could intervene and the prescriptions in Subsections 2a and 2b need
not be modified any further. However, the causality argument seems to
be very reasonable one make (though radiative viscosity
can still work) although one wonders if
the modification of $\nu_p$ and $\nu_m$ as given by Eqn. (7a) and
Eqn. (7c) after multiplying each of them by a factor $(1-v_r^2/v_t^2)^2$
solves the problem. This is because $W_{r\phi}$ in the preshock flow
would be negligible compared to its value in the postshock flow and
the shock would again become `shear-type' and piled up angular
momentum may continuously drive the shock outwards.

\noindent{\bf 3. Evolution of Isothermal Viscous Accretion Disks}

In Chakrabarti \& Molteni (1993) and Molteni \& Sponholz (1994)
the basic procedures for the implementation of the Smoothed Particle
Hydrodynamics in cylindrically symmetric coordinates are presented
and it was shown that the
simulation results agree very well with the theoretical predictions
regarding the shock parameters. In the present paper, we do not discuss
this any more. The only modification over the Chakrabarti \& Molteni
(1993) is the addition of the angular momentum transport equation (1c)
with the possibility to use various dynamical viscosity prescriptions
as discussed in \S 2 above. As before, we use Paczy\'nski
\& Wiita potential (1980) to describe the external Schwarschild geometry.
We wish to note in passing that our code is tested to be free from any
numerical viscosity which is shear type. This is because each of the
pseudo-particles is axially symmetric and interacts as a torus.

\noindent{\large 3a. Test Results}

In Chakrabarti \& Molteni (1993) we already shown that a flow
free from viscosity produces shock waves exactly where shock
waves are predicted in disks. The code we use here has been tested
for adiabatic flows in Molteni and Sponholz (1994).
Here, we first show a test result to indicate that
our code is good enough to produce a Keplerian isothermal disk as well.
Evolution of a ring of matter into a Keplerian disk in presence of
viscosity has been demonstrated long ago (e.g., Pringle, 1981).
In our simulation, we inject particles at the outer edge of the disk
at $r_{o}=100$ with angular momentum $7.00$ which is close to
Keplerian value at $r_o$. The (constant) sound speed
is chosen to be $K=0.005$ and the injection velocity was $v_0=0.003$.
The artificial viscosity parameters were $A=1$ and $B=1$
(see Chakrabarti \& Molteni 1993 where these parameters
are denoted as $\alpha$ and $\beta$). The SPH parameters
were: particle size $h=0.4$ and the particle separation $\delta_p=0.2$
(See, e.g., Monaghan, 1992 for definitions).
The viscosity slowly works on the flow and transports the angular momentum
outwards, which enables matter to fall onto the central black hole. Particles
reaching beyond the outer grid as well as the below $r=1.3$ were removed.
In this way particles with higher angular momentum is removed from the
outer edge and the particles with lower angular momentum is swallowed
by the black hole.
Figs. 1(a-b) shows the angular momentum distribution achieved
after flow becomes steady. Fig. 1a is the result of a simulation
using SS$\alpha$ prescription (Eqn. 4) and Fig. 1b is the result of a
simulation using mass flux balanced viscosity prescription (Eqn. 7d). Alpha
parameters chosen are $\alpha_{s}=\alpha_m=0.25$. In each of these
figures, there are about 500 particles. In both the cases,
final distribution seems to be very close to the Keplerian
distribution (dotted curves) till the marginally stable orbit
at $r=3$ after which the flow falls freely and supersonically to the black
hole. The excellent agreement suggests that the code with the inclusion
of viscosity is working very well.

In all the simulation results listed below, we use the following
quantities: The outer edge is at $r_0=18$, the velocity of
sound $K=0.05$, the specific angular momentum $l=1.89$ and the injection
velocity $v_{ro}=0.1$. Various cases are distinguished by
different viscosity prescriptions.

\noindent {\large 3b. Results using Shakura-Sunyaev Prescription}

In Fig. 2(a-b) we compare Mach number and angular momentum variations
in viscous and inviscid flows. The viscosity parameter
$\alpha_s=0.01$ is chosen everywhere in the viscous disk simulation.
Note that the shock in the viscous disk forms farther out and is weaker
and wider as expected (SKC90b). Fig. 2b shows that a shear shock is formed
with a jump in angular momentum at the shock ($\lambda_+ > \lambda_-$).
This is because the transport rate of angular momentum is higher
in the postshock flows compared to the rate in the preshock flows.
The angular momentum is super-Keplerian in the postshock flow
and almost constant in the preshock flow.

As the $\alpha_s$ parameter is increased, the flow behaviour changes
dramatically. Fig. 3(a-b) shows the evolution of
Mach number and angular momentum when $\alpha_s=0.1$ is chosen.
One notices that the shock rapidly propagates outwards making the
entire flow outside the inner sonic point subsonic, and the angular momentum
distribution resembles more and more Keplerian. Successive curves
are drawn at intervals of $\Delta t=500$. The infall time-scale is $t_{i}=75$
in the same unit. What clearly happens in this case is that the piled
up angular momentum in the post shock flow drives the shock
outwards continuously and no steady solution becomes possible.

The topological properties of a viscous accretion flow are discussed
in detail in SKC90b which we do not repeat here. A salient point that was
observed is that when viscosity is very low, a stable and
an unstable shock may form (in notation of SKC90b, at $X_{s3}$ and $X_{s2}$
respectively) with the stable shock ($X_{s3}$)
gradually becoming weaker as viscosity is increased. When viscosity
crosses a critical value, the stable shock disappears altogether.
Instead, two solutions, both coming from infinity to the horizon
are seemingly allowed, none being suitable for a stable shock formation.
The solution passing through the inner sonic point has a higher dissipation
and entropy (Chakrabarti, 1989, 1990a) and is chosen by the realistic
flow. Our results presented in Fig. 1(a-b) and Fig. 3(a-b) above seem to
correspond to this branch of the solution.

\noindent {\large 3c.  Results using Mass Flux Balanced Viscosity Prescription}

Figure 4(a-b) shows the variation of Mach number and angular momentum
profile for flows evolved with mass flux balanced viscosity prescription
discussed in \S 2 above. There are three curves in each Figure.
The solid curve is obtained with $\alpha_m=0.01$ everywhere and the long
dashed curve is obtained with $\alpha_m=0.01$ only in the subsonic region
while $\alpha_m=10^{-6}$ in the supersonic region.
This latter case was chosen to mimic the flux-limited diffusion
prescription (Narayan, 1992). It is clear that the angular
momentum distribution is roughly monotonic and sub-Keplerian except close
to the black hole where it is super-Keplerian. The introduction of the
flux-limited transport prescription changes the properties
of the flow in obvious ways. The shock becomes stronger, but
still located farther out. The angular momentum distribution
becomes strongly non-monotonic and the shock is of `shear-shock' type
where a jump in angular momentum occurs.

\noindent {\large 3d.  Results using Pressure Balanced Viscosity Prescription}

Figures 5 shows the angular momentum distribution of the disk with
pressure balanced viscosity prescription. Solid curve is obtained
with $\alpha_p=0.1$ everywhere in the flow and the dashed curve
is obtained using $\alpha_p=0.1$ only in the subsonic region
while using $\alpha_p=10^{-5}$ in the supersonic region. The angular momentum
distribution is non-monotonic but is smooth and continuous apart from some
numerical noise. The dashed curve with flux-limited transport prescription
is smoother in comparison. The jump in angular momentum is smaller.
Dotted curve represents the Keplerian distribution for reference purpose.

In Fig. 6, we present a comparison of the angular momentum distributions
obtained using all the three prescriptions discussed above.
The postshock flow shows a significant variation of the distribution, while
the slopes in the supersonic preshock region remain very similar.
Note that $\alpha_p$ is chosen to be an order of magnitude higher, since
the pressure balanced prescription requires the dynamical viscosity
to proportional to the square of the velocities (Eqn. 7a).

\noindent{\bf 4 Concluding Remarks}

In this paper, we have systematically presented numerical evolution
of thin, isothermal accretion disks following various viscosity
prescriptions. We discovered several significant results: we find that
when viscosity is low enough, the shocks are weaker,
wider and form farther out. For high viscosity, the flow does not
produce a stable shock due to qualitative change in the topological property
of the flow. Rather, an unstable shock travels outwards
sweeping the disk and making it subsonic and Keplerian. The flow becomes
supersonic only at the inner sonic point located close to the horizon.
These results agree with the theoretical expectations
(Chakrabarti 1990b). The axisymmetric solutions seem to be stable.
In future, one needs to study if these solutions remain
stable under the non-axisymmetric perturbations as well.

We make here an important observation that the Shakura-Sunyaev
prescription, regarded widely as the working
description of viscosity in accretion disks, does not satisfactorily
describe the shocks in the disks, since the shocks are always shear-type
and the jump in the angular momentum produces a region of a negative slope
(although, we not  see any otherwise unpleasant behaviour in terms
of stability).
In search of an alternate description, it became clear that one could instead
use the pressure balance condition or the mass flux conservation
conditions to define viscous stress. Each of these prescriptions
seems to describe the shocks better. While the pressure
balanced prescription still produces a smaller jump in angular momentum
at the shock, the mass flux balanced prescription gives roughly monotonic,
smooth and continuous distribution of angular momentum in the disk,
including regions across the shock.
When the viscosity is high, this latter prescription reproduces shock-free
Keplerian subsonic solution as well. Thus we believe that
the mass flux balanced prescription is probably more satisfactory.

The flux-limited diffusion condition, when coupled to any of
our prescriptions, produced shear shocks in the disks since the viscous
stress does not remain continuous. It is possible
that in reality, purely hydrodynamical processes are rare and the disk
angular momentum may be transported via magnetic turbulences or radiative
processes. In future, we plan to investigate these issues in more detail.

\newpage

\centerline{\bf REFERENCES}

\noindent

\noindent Chakrabarti, S. K. 1989, ApJ, 347, 365\\
Chakrabarti, S. K. 1990a, {\it Theory of Transonic Astrophysical Flows},
World Scientific Publ. Co. (Singapore)\\
Chakrabarti, S. K. 1990b, MNRAS, 243, 610\\
Chakrabarti, S. K. \& Khanna, R. 1992, MNRAS, 256, 300\\
Chakrabarti, S. K. \& Molteni, D. 1993, ApJ, 417, 671\\
Ferrari, A. Trussoni, E., Rosner, R. \& Tsinganos, K. 1985, ApJ, 294, 397\\
Fukue, J. 1987, PASJ, 39, 309\\
Habbal. S. R.  \& Tsinganos, K. 1983, J. Geophys. Res.. 88, 1965\\
Hawley, J.W., Smarr, L. \& Wilson, J. 1985, ApJ, 277, 296\\
Liang, E. P. T. \& Thompson, K. A. 1980, ApJ, 240, 271\\
Molteni, D., Lanzafame, G. \& Chakrabarti, S. K. 1994, ApJ, (April 10th
issue)\\
Molteni D. \& Sponholz H, Proceedings of the OAT-SISSA International Workshop
on Smoothed Particles Hydrodynamics in Astrophysics 1994, (in press).\\
Monaghan J. J. 1992, Ann. Rev. Astron. Astrophys., 30, 543\\
Narayan, R. 1992, ApJ, 394, 255\\
Paczy\'{n}ski, B. \& Wiita, P. J. 1980, A\&A, 88, 23\\
Pringle, J. E. 1981, Ann. Rev. Astron. Astrophys., 19, 137\\
Shakura, N. I. \& Sunyaev, R. A. 1973, A \& A, 24, 337\\

\newpage

\noindent Fig. 1(a-b): Comparison of the Keplerian angular
momentum distribution (dotted) with the distribution obtained by
numerical simulations (solid) using (a) Shakura-Sunyaev viscosity prescription
($\alpha_s=0.25$) and (b) mass flux balanced viscosity prescription
($\alpha_m=0.25$). Thick solid patches are due to denser number
of particles in some regions.

\noindent Fig. 2(a-b): Comparison of (a) Mach number
variation and (b) angular momentum distribution in
a disk using Shakura-Sunyaev viscosity $\alpha_s=0.01$ (solid curves)
with those in an inviscid disk (dashed curves). The shock in viscous
disk is wider, weaker and farther out. In (b), Keplerian
angular momentum is also shown for comparison. Note the super-Keplerian
region close to the hole and the angular momentum jump at the shock.

\noindent Fig. 3(a-b): Non-steady evolution of the (a) Mach number
variation and (b) the angular momentum distribution in a disk with
a large Shakura-Sunyaev viscosity ($\alpha_s=0.1$). Successive curves
are drawn at intervals of $\Delta t=500$ (infall time scale $t_{i}=75$).
The shock propagating outwards makes the entire flow subsonic except in the
vicinity of the horizon, and the distribution of angular momentum
gradually becomes Keplerian. The steady inviscid disk solution is also
presented for comparison.

\noindent Fig. 4(a-b): Solid curves showing
(a) Mach number and (b) angular momentum distribution of a disk with mass
flux balanced viscosity prescription while long dashed curves
are for flows in which viscosity is suppressed artificially in the supersonic
region following flux-limited diffusion prescription. $\alpha_m=0.1$ was used
in both the cases. Short dashed curves indicate solutions for inviscid flows.
In (b), Keplerian distribution is provided for comparison.

\noindent Fig. 5: Angular momentum distribution of the disk with pressure
balanced viscosity prescription. Solid curve is with $\alpha_p=0.1$
everywhere in the flow and the dashed curve is with $\alpha_p=0.1$ only
in the subsonic region, but $\alpha_p=10^{-5}$ in the supersonic region.
Dotted curve represents the Keplerian distribution for comparison.

\noindent Fig. 6: Comparison of angular momentum distributions in disks
with various viscosity prescriptions: with $\alpha_s=0.01$
(long dashed curve), with $\alpha_m=0.01$ (dotted curve) and
with $\alpha_p=0.1$ (solid, ragged). The marked variation occurs
only in the postshock region. An order of magnitude higher $\alpha_p$
was used, since the dynamical viscosity varies as the square of the velocities.
Keplerian distribution (solid, smooth) is provided for reference.

\end {document}